\documentclass[%
reprint,
amsmath,
amssymb,
aps,
]{revtex4} 

\usepackage{graphicx}
\usepackage{dcolumn}
\usepackage{bm}
\usepackage{empheq}
\usepackage{color}
\usepackage{url}
\usepackage{amsmath, amsthm, amssymb, bm} 
\usepackage{placeins} 
\newcommand{\be}{\begin{equation}}
\newcommand{\ee}{\end{equation}}
\newcommand{\beq}{\begin{equation}}
\newcommand{\eeq}{\end{equation}}
\newcommand{\bea}{\begin{eqnarray}}
\newcommand{\eea}{\end{eqnarray}}

\DeclareMathOperator{\sinc}{\text{sinc}}

\begin{document}


\title{Sensitivity limits of space-based interferometric gravitational wave observatories from the solar wind}


\author{Oliver Jennrich}
\affiliation{European Space Agency, ESTEC, The Netherlands}
\author{Nora Luetzgendorf}
\affiliation{European Space Agency, STScI, USA}
\author{James Ira Thorpe}
\email[]{james.i.thorpe@nasa.gov}
\author{Jacob Slutsky}
\affiliation{NASA Goddard Space Flight Center}
\author{Curt Cutler}
\affiliation{Jet Propulsion Laboratory, California Institute of Technology}
\date{\today}

\begin{abstract}
Space-based interferometric gravitational wave instruments such as the ESA/NASA Laser Interferometer Space Antenna (LISA) observe gravitational waves by measuring changes in the light travel time between widely-separated spacecraft.  One potential noise source for these instruments is interaction with the solar wind, in particular the free electrons in the interplanetary plasma. Variations in the integrated column density of free electrons along the laser links will lead to time-of-flight delays which directly compete with signals produced by gravitational waves. In this paper we present a simplified model of the solar plasma relevant for this problem, anchor key parameters of our model using data from the NASA \emph{Wind}/SWE instrument, and derive estimates for the effect in the LISA measurement. We find that under normal solar conditions, the gravitational-wave sensitivity limit from the free-electron effect is smaller than other noise sources that are expected to limit LISA's sensitivity. 
\end{abstract}


\maketitle


\section{\label{sec:intro}Introduction}
Space-based interferometric gravitational wave instruments will extend the promising new field of gravitational wave astronomy from the high-frequency regime ($10\,\textrm{Hz}\lesssim f_\text{GW} \lesssim 1\,\textrm{kHz}$) probed by current and future terrestrial interferometers to a low-frequency regime ($0.1\,\textrm{mHz}\lesssim f_\text{GW} \lesssim 1\,\textrm{Hz}$) populated with numerous and varied astrophysical sources\cite{GravUniverse}. The ESA/NASA Laser Interferometer Space Antenna (LISA)\cite{2017proposal} will employ a triangular constellation of spacecraft connected by two-way optical links approximately 2.5$\,$Mkm on a side.  The LISA constellation will occupy a heliocentric orbit at 1AU trailing approximately 20$^\circ$ in orbital phase behind the Earth.  As with all bodies in the solar system, the LISA constellation will be bathed in a plasma of charged particles produced by the Sun.  Interaction between this plasma and the laser light traveling along the arms of the LISA constellation will produce fluctuations in the effective optical path length along the arms that will directly compete with similar fluctuations produced by gravitational waves.  The existence of this effect, and its potential to limit sensitivity to gravitational waves, was recognized in early studies of LISA~\cite{LISA_PPA}, where it was estimated to produce length fluctuations with an amplitude spectral density of $5\times10^{-12}\,\textrm{m}\:\textrm{Hz}^{-1/2}$ over the LISA arms, which were 5 Mkm long in the LISA designs of the time. The resulting strain limit of $\sim 10^{-21}\,\textrm{Hz}^{-1/2}$ was sufficiently small to enable the full LISA science case. These estimates were informed by an analysis of long-baseline radio transmissions by Woo \& Armstrong~\cite{WooAndArmstrong}, who estimated fluctuations in the column density of electrons in the solar plasma between Earth and the Viking I \& II spacecraft at Mars. The same effect was recognized as a limiting noise source in searches for micro-Hertz gravitational wave signals using two-way radio links with Cassini at Saturn \cite{TintoAndArmstrong}.  More recently, in-situ measurements of the solar plasma from several spacecraft have provided an opportunity to anchor our models and refine estimates of the effect in LISA. In this paper, we review a simplified model of the solar plasma (\ref{sec:model}), connect this model with data from the SWE instrument onboard the \emph{Wind} spacecraft (\ref{sec:data}), estimate the effect for LISA (\ref{sec:lisa}), and summarize our conclusions (\ref{sec:conclude}).

\section{\label{sec:model}Simplified model of the solar plasma}
The solar wind is a stream of electrons, protons, and heavier charged particles that originate in the Sun and propagate at high speeds into interplanetary space.  At low solar latitudes, the typical velocity of the solar wind is $V\sim400\,\textrm{km}/\textrm{s}$.  In the rest frame of the wind, the particles and magnetic field are turbulent, meaning that on the length scales of interest to LISA,  fluctuations in relevant quantities such as particle density are characterized by a Kolomogorov spectrum of the form $S(k)dk\propto k^{-5/3}dk$, where $S(k)$ is the power spectral density of the fluctuations parametrized by their wavenumber $k$.  For the case of propagation along the LISA optical links, the particular quantity of interest is the electron number density, which can be written as (see, e.g., Chapt. 15 of \cite{BlandfordThorne}):
\begin{equation}
\big\langle\delta \tilde n_e^*(\vec k) \ \delta \tilde n_e(\vec k') \big\rangle \equiv P_0 k_0^{11/3}  |\vec k|^{-11/3} (2\pi)^3 \delta(\vec k - \vec k')
\label{eq:ne3d}
\end{equation}
where $\delta n_e(\vec x)$ are the fluctuations of the electron number density in the spatial coordinates defined by $\vec x$, $\delta\tilde{n}_e(\vec k)$ is the Fourier transform of those fluctuations parametrized by the 3-D wavenumber $\vec{k}$ conjugate to $\vec{x}$, $P_0$ describes the overall amplitude of the number density fluctuations, and $k_0$ is a reference wavenumber at which $P_0$ is defined.


The expression for the spectrum of three-dimensional Kolmogorov fluctuations in (\ref{eq:ne3d}) can be used to derive an expression for the spectrum of fluctuations along a single axis, $\delta N_e\equiv\delta n_e(0,0,z)$ which we will show shortly is related to both the interpretation of in-situ electron density measurements from spacecraft as well as the interaction between the LISA optical links and the solar electron plasma. We begin by separating those fluctuations into their components along the $z$-axis, measured by 1D wavenumber $k_z$, and those perpendicular to that axis, measured by  wavenumber $k_\perp$.  We then use the expression in (\ref{eq:ne3d}) and integrate using the expression $d^3\vec k = 2\pi k_\perp d k_z dk_\perp$:
\begin{widetext}
\bea
 \bigg\langle \delta \tilde {N_e}^*(k_z) \ \, \delta \tilde N_e(k'_z)  \bigg\rangle \, &
 =  & \bigg\langle \int{dz} \, e^{-i k_z z}\, \int{dz'} e^{i\,{k'_z} z'}
\int \frac{d^3\vec m}{(2\pi)^3} \, \delta \tilde n^*(\vec m)\ e^{i\,m_z z}\,  \frac{d^3\vec m'}{(2\pi)^3}\,  \delta \tilde n(\vec m') \, e^{-i\,m'_z z'}\, \bigg\rangle \nonumber \\
& = & P_0 k_0^{11/3} \delta(k_z - k'_z) \, \int_0^{\infty} (m^2_{\perp} + k^2_z)^{-11/6}  m_\perp dm_\perp \nonumber \\
& = & \frac{3}{5} \,P_0\, k_0^{11/3}\,  {k_z}^{-5/3} \delta(k_z - k'_z)   
\label{eq:ne1d}
\eea
\end{widetext}

For reference, units of various quantities are as follows:
$n_e(\vec r)$,  $\delta(\vec k - \vec k')$, and $P_0$ all have units cm$^{-3}$;
$\tilde n_e(\vec k)$ has units cm$^0$; $\delta N_e(k_z)$ has units cm$^{-2}$; and it
follows that both sides of  Eq.~(\ref{eq:ne1d}) have units cm$^{-4}$.
 
\section{\label{sec:data}Solar plasma measurements from Wind/SWE}
\begin{figure}[htbp]
\begin{center}
\includegraphics[width=20cm]{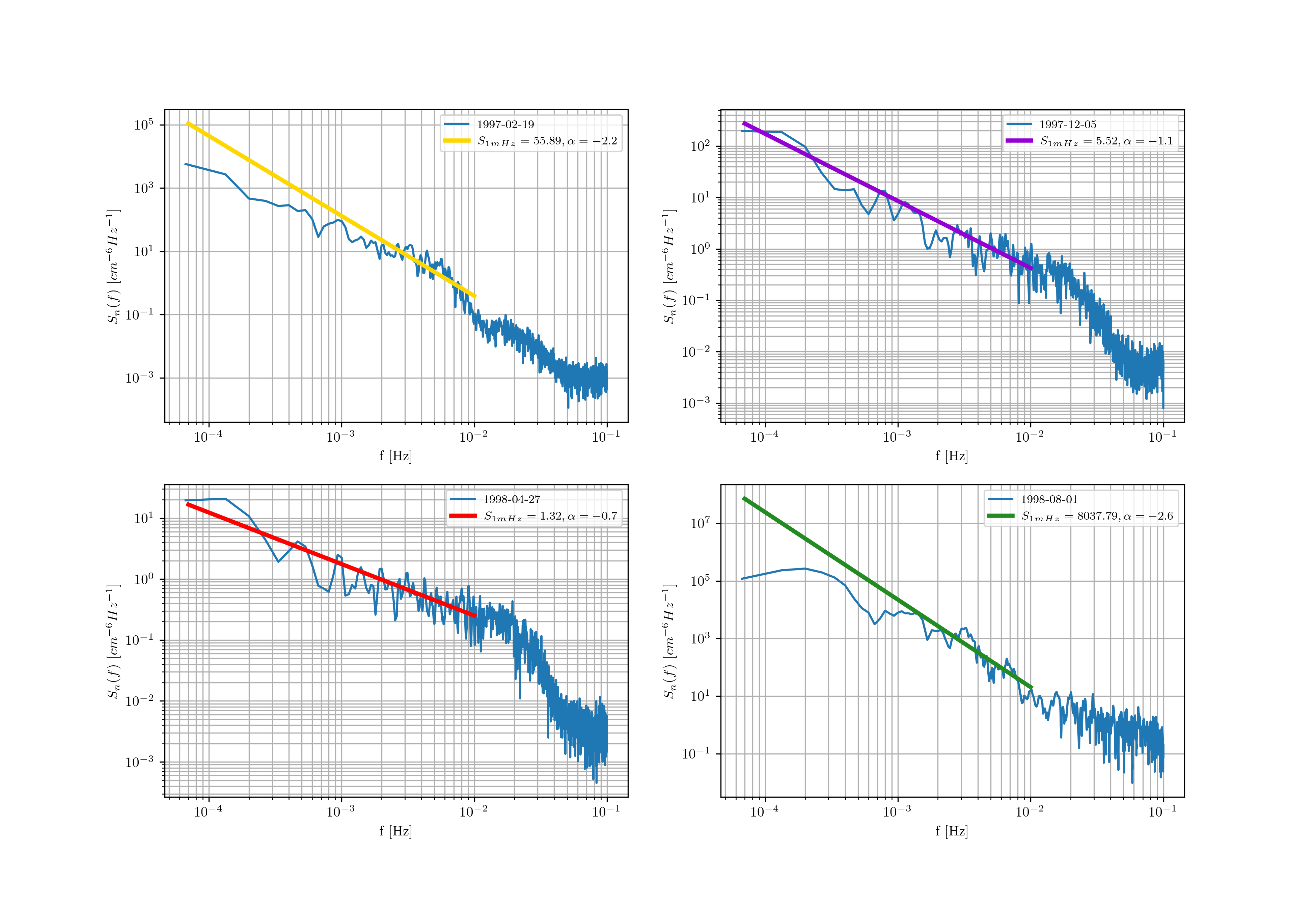}
\caption{Example spectra of electron density measurements from \emph{Wind}/SWE and fits to a simple power-law $S_{N_e}=S_{\text{1\,mHz}}(f/1\,\textrm{mHz})^\alpha$ in the region $10^{-4}\,\textrm{Hz} < f < 10^{-2}\,\textrm{Hz}$. Each spectra is made with a 24-hr period of data. For the ensemble of data considered here, the four panels represent (clockwise from top left) the spectra with the median $S_{\text{1\,mHz}}$, median time sample, minimum $S_{\text{1\,mHz}}$, and maximum $S_{\text{1\,mHz}}$.}
\label{fig:WINDSpec}
\label{default}
\end{center}
\end{figure}

The \emph{Wind} spacecraft is a NASA heliophysics mission launched in 1994 to study the solar plasma and its interaction with the Earth's magenetosphere. One of \emph{Wind's} instruments is the Solar Wind Experiment (SWE)
\cite{Ogilvie1995} from which a calibrated measure of electron density is derived.  \emph{Wind's} operations have consisted of two main phases: a ``near-Earth" phase (1994--2004) when the spacecraft made a series of trips through the Earth's magentosphere and an ``L1" phase (2004--present) when the spacecraft occupied a Lissajous orbit around the 1st Sun-Earth Lagrange point, approximately 1.5$\,$Mkm in the Sunward direction from Earth.  In both cases, the motion of the spacecraft in the rest frame of the local solar wind is an approximately constant velocity of $V\approx 400\,$km/s in the Sunward direction. This velocity is large compared with the velocity scales of the turbulent motion (e.g. the spacecraft's motion through the wind is supersonic) and also at least an order of magnitude larger than the orbital velocity of the spacecraft relative to the Sun. Under both of these approximations, a spacecraft like \emph{Wind} will measure a set of turbulent electron density fluctuations that are `frozen' into the plasma and carried across the detector by the bulk velocity of the solar wind, sometimes referred to as the Taylor hypothesis\cite{chen_2016}. This allows us to relate the in-situ measurements of electron density made by \emph{Wind}/SWE to the expressions for electron density introduced in section \ref{sec:model}.  Specifically, the timeseries of measured electron density fluctuations is directly related to the spatial fluctuations along the $z-$axis defined by the flow of the solar wind, $\delta N_e(t)\equiv \delta n_e(0,0,-V t)$.  The power spectrum of $n_e(t)$ at an in-situ spacecraft is then easily calculated from (\ref{eq:ne1d}):
\be\label{eq:P_omega}
 \bigg\langle \delta \tilde {N_e}^*(\omega) \ \, \delta \tilde N_e(\omega')  \bigg\rangle =  \frac{3}{5} \,P_0\, k_0^{11/3}\,  V^{2/3} {\omega}^{-5/3} \delta(\omega - \omega')
 \ee
which we can re-express in terms of the spectral density in frequency space as
\be
S_{N_e}(f) =   \frac{4\pi}{(2\pi)^{5/3}}\frac{3}{5} \,P_0\, k_0^{11/3}\, V^{2/3} f^{-5/3}\, ,
\label{eq:ne_f}
\ee
where $f\equiv\omega/2\pi$ is the cycle frequency of the fluctuations, and where, every time we write $S(f)$ in this paper, we are using the single-sided convention.
%
To verify this simple model of electron density fluctuations, we selected an ensemble of data from \emph{Wind}/SWE between January 1997 and November of 1998, corresponding to times when the SWE instrument was operating in its nominal configuration (prior to an anomaly that occurred in 2002 \cite{Wilson2021}) and when the spacecraft was at least 10 Earth radii from Earth, placing it sufficiently outside the magnetosphere so as to measure a representative interplanetary environment. For each of the 620 days with available electron density timeseries data meeting our criteria, we computed the spectrum of their fluctuations using the following algorithm\footnote{A Jupyter notebook used to produce the analysis in this paper is available at https://github.com/ithorpe/solarWindLISA}: resample the timeseries to a regular 5-s sampling grid, remove single-sample outliers, and compute the power spectral density using Welch's method of overlapped averaged periodograms.  Figure \ref{fig:WINDSpec} shows example spectra of electron density fluctuations for four different dates, along with the best fit to a simple power-law model $S_{N_e}=S_{\text{1\,mHz}}(f/1\,\textrm{mHz})^\alpha$ over the range $10^{-4}\,\textrm{Hz}< f < 10^{-2}\,\textrm{Hz}$.  Clockwise from top left, the four spectra correspond to the dates with the median PSD at 1mHz (1997-02-19), the mid-point in ensemble (1997-12-05), the minimum PSD at 1mHz (1998-04-27), and the maximum PSD at 1mHz (1998-08-01).

\begin{figure}[htbp]
\begin{center}
\includegraphics[width=12cm]{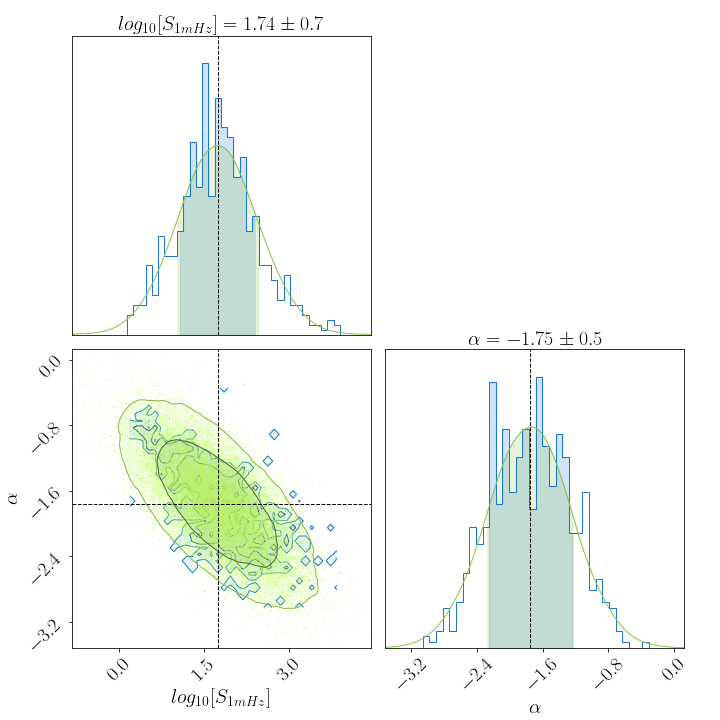}
\caption{Distribution of power-law fit parameters $S_{\text{1\,mHz}}$ and $\alpha$ for the 620 samples in the ensemble. A multivariate normal distribution with the same covariance is over-plotted in green.}
\label{fig:WINDSpecFit}
\label{default}
\end{center}
\end{figure}

Figure \ref{fig:WINDSpecFit} shows the distribution of power-law fitting parameters $S_{\text{1\,mHz}}$ and $\alpha$  for all 620 samples in the ensemble. Overplotted in green is a multi-variate normal distribution corresponding to $\log_{10}(S_{\text{1\,mHz}})\,=\,1.74\pm0.7$ and $\alpha=-1.75\pm0.5$.  A clear anti-correlation between spectral amplitude and spectral index is observed, with a measured statistical correlation of -0.7. Note that the median spectral index agrees well with the value of $-5/3$ predicted in (\ref{eq:ne_f}).  

\begin{figure}[htbp]
\begin{center}
\includegraphics[width=20cm]{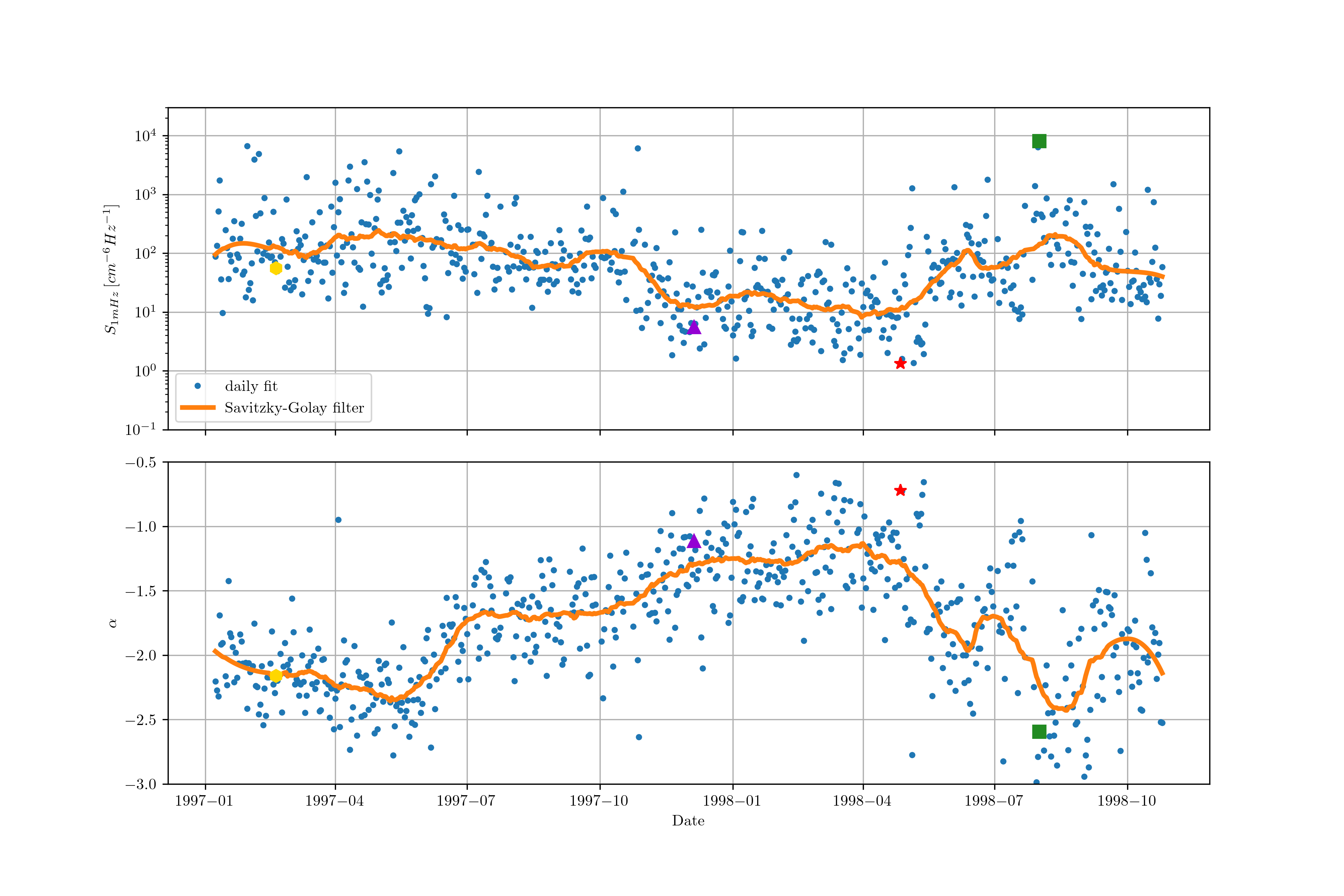}
\caption{Time-variation of the  power-law fitting parameters $S_{\text{1\,mHz}}$ and $\alpha$  for all 620 samples in the ensemble of \emph{Wind}/SWE data. Points are parameters for daily fits of the spectra, trendline is a 71-point, 3rd-order Savitzky-Golay filter, and markers correspond to the four examples from Figure \ref{fig:WINDSpec}. Top panel shows the evolution of power-law amplitude $S_{\text{1\,mHz}}$ and bottom panel shows evolution of spectral index $\alpha$.}
\label{fig:WINDFitTime}
\label{default}
\end{center}
\end{figure}

Figure \ref{fig:WINDFitTime} shows the time-variation of the  power-law fitting parameters $S_{\text{1\,mHz}}$ and $\alpha$  for all 620 samples in the ensemble. The four samples from Figure \ref{fig:WINDSpec} are indicated using the colored markers and as well as a trendline produced by applying a 71-point 3rd-order Savitzky-Golay filter \cite{SavitzkyGolay_1964} to the data.
Over the approximately two years of data, the power of the fluctuations varies by about 1.5 orders of magnitude. The anti-correlation between fluctuation amplitude and spectral index is also readily apparent.  For context, the time period when this data was taken during a period of low solar activity (as measured by number of sunspots) during the beginning of solar cycle 23\cite{NOAAsunspots}. Electron density is known to be anti-correlated with sunspot number \cite{ISSAUTIER20052141}, thus the results shown here correspond roughly to an epoch of \emph{maximum} electron density within the solar cycle.  If the LISA schedule and solar activity adhere to current predictions, science operations will begin during a similar portion of solar cycle 26.

\section{\label{sec:lisa}Estimated effect for LISA}
We now turn our attention to the effect the solar plasma will have on the optical interferometric measurement made between pairs of LISA spacecraft. The spacecraft are arranged in a triangular constellation approximately 2.5 Mkm on a side that is placed in an Earth-like heliocentric orbit, lagging behind (or potentially leading ahead) of the Earth by approximately 20$^\circ$ in orbital phase, equivalent to an Earth-constellation distance of approximately 50 Mkm.  The plane of the constellation is inclined with respect to the ecliptic by 60$^\circ$.
\\
As the optical beams traverse the space between the satellites, they will interact with the free electrons in the solar wind, resulting in a phase shift that will be present in the LISA measurement.  This phase shift or time delay can be modeled as an effective index of refraction,
\bea
\mu & = & \sqrt{1-\left(\frac{\omega_p}{\omega}\right)^2} \nonumber \\
& \approx &1-\frac{1}{2}\left(\frac{\omega_p \lambda}{2\pi c}\right)^2 \label{eq:mu}
\eea
where $\mu$ is the index of refraction, $\lambda$ is the laser wavelength, $c$ is the speed of light, and $\omega_p$ is the plasma frequency for the solar wind electrons, given by:
\bea
\omega_p^2 & = & \frac{4\pi\,  n_e(t,\vec{r}) e^2}{m_e} \nonumber \\
& = & \left(5.64\times 10^{4}\,\textrm{s}^{-1}\right)^2 \times \left( \frac{n_e(t,\vec{r})}{\textrm{cm}^{-3}}\right)\label{eq:omp}
\eea
where $e$ is the fundamental electric charge, $m_e$ is the mass of the electron, and $n_e$ is the electron density introduced previously. (Note we are using cgs units, with charge measured in statCoulombs.) The relevant quantities for LISA are the fluctuations in the optical path length along the arms caused by variations in the electron density. In the following, we make an estimate for the magnitude of those fluctuations based on the simple model of the turbulent solar wind described in section \ref{sec:model}.  We do this first for a single LISA arm, which is useful for making rough comparisons with other noise sources in the LISA measurement, and then for differential fluctuations in a two-arm LISA observable, which represents a more accurate estimate of the effect.

\subsection{\label{sec:one_arm}The single-arm case}
We begin by defining a coordinate system with origin at the center of LISA, the $\hat z$ direction pointing towards the Sun, the $\hat x$ direction normal to ecliptic, the $\hat y$ direction lying in the ecliptic, and the LISA arm pointing along the unit vector $\hat L_1$. The optical path length fluctuations along the LISA arm caused by electron density fluctuations are then given by

\be
\delta L_1(t) = \chi \int_0^L \delta n_e( \vec r = l \hat L_1 - V t \hat z)\, dl 
\label{eq:L1arm}
\ee
where:
\be
\chi\equiv\frac{\lambda^2 e^2}{2\pi m_e c^2}\approx5.08\times10^{-22}\,\textrm{cm}^3.
\label{eq:chi}
\ee

\begin{figure}[htbp]
	\begin{center}
		\includegraphics[width=14cm]{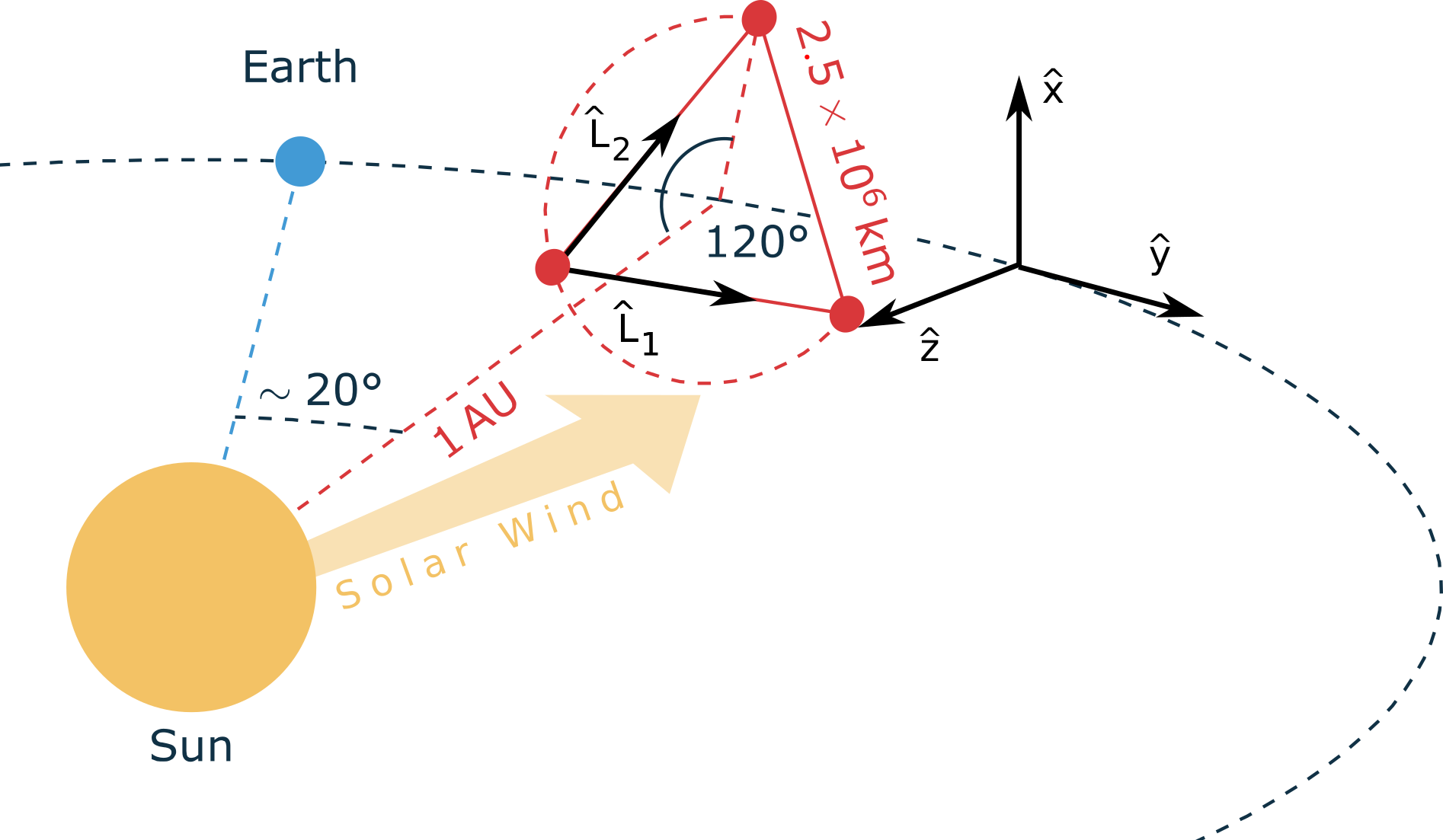}
		\caption{Geometry of the LISA arm constellation with respect to the direction of the solar wind.}
		\label{fig:geo}
		\label{default}
	\end{center}
\end{figure}

\begin{figure}[htbp]
\begin{center}
\includegraphics[width=14cm]{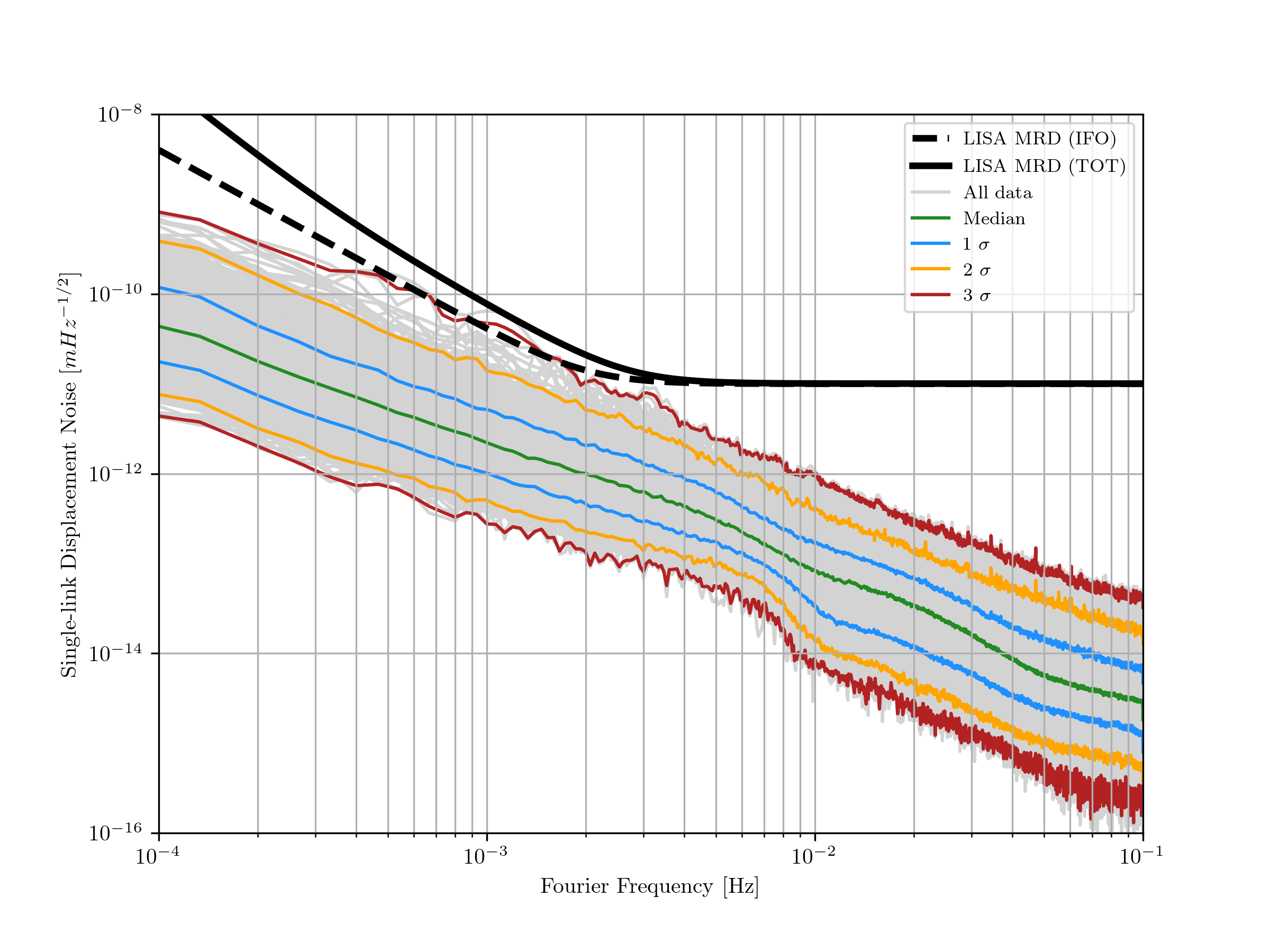}
\caption{Estimated single-link displacement noise for LISA based on Wind/SWE measurements of electron density spectra and the transfer function in (\ref{eq:L1arm_approx}). The gray traces represent each of the 572 daily spectra as described in section and the green trace represents the median at each frequency bin. The blue, orange, and red traces represent the 1-, 2-, and 3-$\sigma$ intervals for each bin respectively.  The two thick black lines represent LISA single-link sensitivity requirements from the LISA Mission Requirements Document, the dashed line represents LISA's sensitivity limit from the interferometric measurement system and the solid line represents the combined limit of the interferometer and the test-mass motion.}
\label{fig:1arm}
\label{default}
\end{center}
\end{figure}

Each LISA arm points in a direction that has both a tangential and a radial component in the heliocentric coordinate system. Integration of (\ref{eq:L1arm}) along the solar wind (radial) direction produces a fundamentally different result than a tangential direction.  For LISA, this mix of radial and tangential components will evolve as the constellation undergoes its orbit around the Sun. Appendix \ref{sec:onearm} provides a detailed derivation of the apparent optical path length induced by electron plasma with the following result:
\be
\big\langle \tilde{\delta L^*_1}(\omega) \tilde{\delta L_1}(\omega') \big\rangle\approx(L \chi)^{2}\left[\frac{3}{5} \,P_0\, k_0^{11/3}\, V^{2/3} |\omega|^{-5/3}\right]\left(\frac{25}{9}\right){\beta_1}^{5/3}  \left(\frac{V}{L\omega}\right) \delta(\omega - {\omega}') 
\label{eq:L1arm_approx}
\ee
where $\frac{\sqrt{3}}{2}  \leq \beta_1 \leq 1$ is an angle that represents the time-evolving orientation of the LISA constellation. The term in square brackets is the power spectrum of in-situ electron density measurements from (\ref{eq:ne_f}), which allows the LISA effect to be estimated from \emph{Wind}/SWE data using a simple transfer function:
\be
S_{L1}(f)\approx(L \chi)^{2}\left(\frac{25}{9}\right)\ {\beta_1}^{5/3}  \left(\frac{V}{2\pi L f}\right)S_{\text{Ne}}(f)
\label{eq:simpleTF}
\ee
It is insightful to break the transfer function into parts: it is   $(L \chi)^{2}$, times a factor of order one ($\frac{25}{9} \approx 2.78$), times the geometrical factor  ${\beta_1}^{5/3} $ (which varies between 0.79 and 1.0 as LISA's orientation changes), times the quantity $(\frac{V}{2\pi f L})$, which comes from the partial "averaging out" of fluctuations along the arm.  For $f = 5\,\text{mHz}$ and $V = 400\,\text{km/s}$, $V/(2\pi f L) = 5.1\times 10^{-3}$.  Figure \ref{fig:1arm} shows the spectra obtained from \emph{Wind}/SWE multiplied by the transfer function in (\ref{eq:simpleTF}) for the case where $\beta=1$.  Since the observed spectra exhibit some non-stationarity, all spectra are plotted with the median, 1-, 2-, and 3-$\sigma$ values for each frequency bin shown.  For comparison, the single-link equivalent displacement noise requirements from the LISA Mission Requirements Document (MRD) \cite{MRD_LISA} are plotted as thick black lines. The dashed line shows the requirement on the LISA interferometry system alone whereas the dashed curve includes the additional limit on sensitivity that arises from imperfections in the free-fall of the LISA test masses. 

\subsection{\label{sec:two_arm}Differential arms, TDI, and other effects}
While the LISA MRD specifies sensitivity in terms of equivalent single-link displacement, LISA is fundamentally a multiple-arm interferometer.  A differential measurement between pairs of LISA arms allows suppression of laser frequency noise which would otherwise overwhelm both the gravitational wave signal as well as other interferometer noise sources by 7-8 orders of magnitude. To evaluate the effect of  solar electron plasma effects on the differential measurement, the analysis from Appendix \ref{sec:onearm} can be easily extended.  We first introduce a second arm defined by
\be
\hat L_2 = \alpha_2 \hat z + \beta_2 \hat L_{2,\perp}  \label{eq:L2hat}
\ee
where $\alpha_2$ and $\beta_2$ are the analogous direction angles for the second arm.  The differential length fluctuations $\delta L$ can then be expanded into the length fluctuations in the individual arms plus cross-terms that represent the correlation between the arms: 
\be\label{expand}
\,\big\langle \tilde{\delta L^*}(\omega) \tilde{\delta L}(\omega') \big\rangle  = \,\big\langle \tilde{\delta L^*_1}(\omega) \tilde{\delta L_1}(\omega') \big\rangle 
+ \,\big\langle \tilde{\delta L^*_2}(\omega) \tilde{\delta L_2}(\omega') \big\rangle \\   
- \,\big\langle \tilde{\delta L^*_1}(\omega) \tilde{\delta L_2}(\omega') \big\rangle  - \,\big\langle \tilde{\delta L^*_2}(\omega) \tilde{\delta L_1}(\omega') \big\rangle
\ee

The first two terms are described by (\ref{eq:L1arm_approx}) and are identical except that $\beta_1 \rightarrow \beta_2$ for the second term. Appendix \ref{sec:twoarm} shows that the cross-terms can be neglected with the result that the power spectrum of electron density fluctuations for the differential case is approximately twice that of the power spectrum for the single arm case. Since the noise requirements for differential measurements are similarly higher, the relative importance of the effect is unchanged

While the analysis above represents a simple differential measurement, LISA's approach is somewhat different. Since the individual arm lengths in the constellation differ by up to $\sim0.5\%$ and are slowly varying over the mission lifetime, a simple differential measurement would not sufficiently suppress the laser frequency noise present in the interferometer. Instead, a technique known as Time-Delay-Interferometry (TDI) (see, e.g. \cite{2021LRR....24....1T}) is applied to combine multiple measurements of the single-link interferometers at different time epochs to generate observables that further suppress laser frequency noise while retaining GW signals.  The TDI combinations are designed to exploit correlations in the laser frequency noise terms at different points and times within the constellation. Noise sources which do not have these same correlations, such as the shot noise and test mass acceleration noise that make up the single-link displacement curves in Figure \ref{fig:1arm} are not suppressed by TDI.  Since the analysis in Appendix \ref{sec:twoarm} demonstrates that the correlation of electron density noise between multiple arms is negligible, TDI will have no effect on the \emph{relative} impact of the electron density noise to other uncorrelated noise sources such as photon shot noise in terms of GW sensitivity.  

\section{\label{sec:conclude}Discussion}
Optical path length fluctuations induced by time-varying electron plasma densities in the solar wind are a potentially important noise source for space-based gravitational wave interferometers such as LISA. We have presented here a simple model of the electron density spectrum, demonstrated that it matches reasonably well with data from the \emph{Wind}/SWE instrument, and estimated the resulting effect for LISA in both simple one-arm models and accounting for the more complex geometry of the constellation.  Our revised estimate is a factor of $\sim 2$ lower than the estimate in earlier LISA studies \cite{LISA_PPA}. It is furthermore consistent with estimates from long-baseline radio measurements\cite{WooAndArmstrong}.  Our conclusion is that the resulting effect for LISA is not expected to limit sensitivity to gravitational wave sources except possibly during rare solar events, when spacecraft operations may be impacted in other ways (e.g. through test mass charging). In any event, rare interruptions of the link from such events would be entirely consistent with the anticipated duty cycles of instruments like LISA which are expected to be around 75\%.  It is clearly a noise source which merits continued attention by the designers of LISA and other future space-based interferometers such as TianQin, Taiji, etc. Indeed, a recent analysis of the effect for the proposed geocentric TianQin mission made using modern MHD simulations of the near-Earth solar plasma found qualitatively similar results \textemdash  the sensitivity of TianQin to gravitational waves will not be limited by solar plasma effects, but their contribution to the overall noise budget is not negligible\cite{TianQinPlasma}.  For second-generation space missions with increased displacement sensitivity, this may be a noise source that requires mitigation of some kind. Possibilities for mitigation include in-situ measurements of the electron density that would allow for post-processing subtraction of the effect or multi-frequency laser systems that would allow the electron density fluctuations along the arm to be directly measured and subtracted.

\appendix

\section{Optical pathlength fluctuations in a 1-D kolmogorov spectrum}

\subsection{\label{sec:onearm}The single arm case}
We begin by defining a coordinate system with origin at the center of LISA, the $\hat z$ direction pointing towards the Sun, the $\hat x$ direction normal to ecliptic, the $\hat y$ direction lying in the ecliptic, and the LISA arm pointing along the unit vector $\hat L_1$ (see Figure \ref{fig:geo}), which is in turn represented in terms of angles $\alpha_1$ and $\beta_1$ as follows:
\be
\hat L_1 = \alpha_1 \hat z + \beta_1 \hat L_{1,\perp}  \label{eq:Lhat}
\ee
where $\hat L_{1,\perp}$ is a unit vector orthogonal to $\hat z$.  It is worth noting that the geometry of the LISA orbits require the values of $\alpha_1$ and $\beta_1$ to be determined by a single angle that gives the orientation of the constellation within its plane and that and that $-\frac{1}{2}  \leq \alpha_1 \leq \frac{1}{2}$ and 
$\frac{\sqrt{3}}{2}  \leq \beta_1 \leq 1$.

The optical path length fluctuations along the LISA arm caused by electron density fluctuations are then given by
\be
\delta L_1(t) = \chi \int_0^L \delta n_e( \vec r = l \hat L_1 - V t \hat z)\, dl 
\label{eq:L1arm_appendix}
\ee

To connect with the model of the electron density from section \ref{sec:model}, we expand $n_e(\vec r)$ into its spatial Fourier components, 
\be
\delta L_1(t)= \chi \int_0^L dl \int \frac{d^3\vec k}{(2\pi)^3} \delta \tilde n_e(\vec k) e^{i \vec k \cdot (l \hat L _1- V t \hat z)}\, dl
\ee

Expanding  $\hat L_1$ using (\ref{eq:Lhat}) and taking the Fourier transform of $\delta L_1(t)$, we find
\bea
\tilde{\delta L_1}(\omega) &=& \chi \int_0^L dl \int \frac{d^3\vec k}{(2\pi)^3} \delta \tilde n_e(\vec k) e^{i l(\alpha_1 k_z  + \beta_1 \vec k_{\perp}\cdot \hat L_{\perp} )} 2\pi \delta(\omega - V k_z) \\
 &=&\chi \int_0^L dl \int \frac{d^3\vec k}{(2\pi)^3} \delta \tilde n_e(\vec k) e^{i l(\alpha_1\omega/V  + \beta_1 \vec k_{\perp}\cdot \hat L_{\perp} )}2\pi \delta(\omega - V k_z)\\
&=& \chi \int \frac{d^3\vec k}{(2\pi)^3} \delta \tilde n_e(\vec k) \frac{1}{i\gamma_1} \big[e^{i \gamma_1 L}  \ -1 \big] 2\pi \delta(\omega - V k_z) \, ,
\eea

where we have defined
\be
\gamma_1 \equiv  \alpha_1 \omega/V  + \beta_1 \vec k_{1,\perp}\cdot \hat L_{1,\perp} \, .
\ee
The above implies that
\be\label{L}
\big\langle \tilde{\delta L^*_1}(\omega) \tilde{\delta L_1}(\omega') \big\rangle = \chi^{2} \int \frac{d^3\vec k}{(2\pi)} \frac{d^3\vec k'}{(2\pi)^3} \big\langle \delta \tilde n^*_e(\vec k) \ \delta \tilde n_e(\vec k')\big\rangle 2\delta(\omega - V k_z) \delta(\omega - V k'_z) C C^*
\ee
where 
\be
C \equiv \frac{1}{i\gamma_1} \big[e^{i \gamma_1 L}  \ -1 \big] 
\ee
Next we use the Kolmogorov spectrum for the electron density fluctuations from (\ref{eq:ne3d}) and integrate over $\vec k'$ and $k_z$ to get

\be\label{LL1}
\big\langle \tilde{\delta L^*_1}(\omega) \tilde{\delta L_1}(\omega') \big\rangle = \frac{\chi^{2} P_0 k_0^{11/3}}{V}\int \frac{d^2\vec k_{\perp}}{(2\pi)^2} \big( (\omega/V)^2 + k_{\perp}^2\big)^{-11/6}  (2\pi) \delta(\omega - \omega') C C^*
\ee
where we have used that
\be
\int d k_z \delta(\omega - V k_z) \delta(\omega' - V k_z) = V^{-1} \delta(\omega - \omega') \, .
\ee
It is straightforward to show that $C C^*$ can be expressed as
\be\label{CCstar}
CC^* = L^2 \sinc^2(\gamma_1 L/2) \, ,
\ee
where $\sinc(x) \equiv x^{-1} \sin(x)$. Then the right-hand side of Eq.~(\ref{LL1}) becomes

\be\label{term1}
 \left (\chi L\right)^{2}  \frac{P_0 k_0^{11/3}}{V} \int \frac{d^2\vec k_{\perp}}{2\pi}  \sinc^2\frac{\gamma_1 L}{2}\, \delta(\omega - {\omega}') \left( \left(\frac{\omega}{V}\right)^2 + k_{\perp}^2\right)^{-11/6}
 \ee

Next we do the following coordinate transformation.  First we shift the origin of the $\vec k_{1,\perp}$ plane to the point 
$\vec k_{1,\perp} = -(\alpha_1/\beta_1)(\omega/V) \hat L_{1,\perp}$ (noting that $\gamma_1$ vanishes at the new origin), and 
starting at this new origin introduce the orthonormal coordinate system $(k_u, k_w)$, where $\hat k_u = \hat L_{1,\perp}$ and 
$\hat k_w$ is the unit vector orthogonal to $\hat k_u$.  Next note that $\gamma_1 = \beta_1 k_u$ (and so is independent of $k_w$)
and that the $k_{1,\perp}^2$ in (\ref{term1}) is given by
\be
k_{1,\perp}^2 = \left(k_u - \frac{\alpha_1 \omega}{\beta_1 V}\right)^2 + k_w^2 
\ee
Given the above, we can re-write (\ref{term1}) as
\be\label{term2}
(\chi L)^{2}  \frac{P_0 k_0^{11/3}}{V}  \int \frac{dk_{u} dk_w}{2\pi} \sinc^2\frac{k_u L}{2}\, \delta(\omega - \omega')\left( \left(\frac{\omega}{V}\right)^2 + \left(k_u - \frac{\alpha_1 \omega}{\beta_1 V}\right)^2 + k_w^2\right)^{-11/6} 
\ee
It is simplest to do the integral over $k_w$ first.  Using that
\be
\int_{-\infty}^{\infty} (A + x^2)^{-11/6}dx = 2 A^{-4/3} \int_{0}^{\infty} (1 + x^2)^{-11/6}dx \approx \frac{5}{3} A^{-4/3}
\ee
(where the last approximation is good to about $1\%$), (\ref{term1}) becomes
\bea
& &\frac{5}{3} \left(\chi L\right)^{2}  \frac{P_0 k_0^{11/3}}{V} \,  \int \frac{dk_u }{2\pi} \sinc^2\frac{\beta_1 k_u L}{2} \,\delta(\omega - \omega')\left( \left(\frac{\omega}{V}\right)^2 + \left(k_u - \frac{\alpha_1 \omega}{\beta_1 V}\right)^2\right)^{-4/3} \\
&=& \frac{5}{3} \left(\chi L\right)^{2}  \frac{P_0 k_0^{11/3}}{V} \,  L^{8/3} \int \frac{dk_u }{2\pi} \sinc^2\frac{\beta_1 k_u L}{2} \,\delta(\omega - \omega')\left( \left(\frac{L\omega}{V}\right)^2 + \left(k_u L - \frac{\alpha_1}{\beta_1}\frac{L\omega}{V}\right)^2\right)^{-4/3} \\
&=& \frac{5}{3} \left(\chi L\right)^{2}  \frac{P_0 k_0^{11/3}}{V} \,  L^{8/3} \left( \frac{ L\omega}{V}\right)^{-8/3}\int \frac{dk_u }{2\pi}  \sinc^2\frac{\beta_1 k_u L}{2}\, \delta(\omega - \omega')
\left( 1 + \left(\frac{k_u V}{\omega} - \frac{\alpha_1}{\beta_1}\right)^2\right)^{-4/3}
\eea
Because of the factor $\sinc^2(\beta_1 k_u L/2)$, the integral will be dominated by the contributions in the region $|k_u| \lesssim 2/L$.
We are most interested in the regime $\omega \gg V/L$, or $f > 2.55 \times 10^{-5}\,\mathrm{Hz}$  (for $V = 400\,\mathrm{km/s}$ and $L = 2.5\times10^6\,\mathrm{km}$), where we can neglect the term $k_u V/\omega$ in the expression
$\left( 1 + (k_u V/\omega - \frac{\alpha_1}{\beta_1})^2\right)$ in the last line above, which then becomes
\bea
& &\frac{5}{3} \left(\chi L\right)^{2}  \frac{P_0 k_0^{11/3}}{V} \,  L^{8/3}\left(\frac{L\omega}{V}\right)^{-8/3} \left( 1 + \left(\frac{\alpha_1}{\beta_1}\right)^2\right)^{-4/3}
\int \frac{dk_u }{2\pi}  \sinc^2\frac{\beta_1 k_u L}{2}\, \delta(\omega - \omega') \\
&=&\frac{5}{3} \left(\chi L\right)^{2}  \frac{P_0 k_0^{11/3}}{V} \, L^{8/3}\left(\frac{L\omega}{V}\right)^{-8/3} \left( 1 + \left(\frac{\alpha_1}{\beta_1}\right)^2\right)^{-4/3}
\frac{1}{\beta_1 L}\, \delta(\omega - \omega') \\
&=& \left(\chi L\right)^{2}\,\left[\frac{3}{5} P_0\, k_0^{11/3}\, V^{2/3} |\omega|^{-5/3}\right]\,\frac{25}{9} \, {\beta_1}^{5/3}  \,\frac{V}{L\omega} \,\delta(\omega - {\omega}') \label{final} ,
\eea

where in the second line we use the integral:

\be
\int_{-\infty}^{\infty} dk_u \sinc^2\frac{\beta_1 k_u L}{2} = \frac{2 \pi}{\beta_1 L} \, .
\ee

and in the last line we we used the fact that  ${\beta_1}^{-1} \big( 1 + (\frac{\alpha_1}{\beta_1})^2\big)^{-4/3} $ simplifies to  ${\beta_1}^{5/3}$. The term in square brackets in \eqref{final} is the expression from \eqref{eq:ne_f} for the power spectral density of electron density fluctuations measured by a "stationary" in-situ spacecraft. The remaining terms in \eqref{final} can thus be viewed as a transfer function between that measurement and the effect in a LISA arm.
\\

The integral for $\delta L_1$ is dominated by a thin elliptical region in the $\vec k_{1,\perp}$ plane.  The center of the ellipse is the point we defined as $\vec k_{1,\perp}$,  its major axis lies along the line where
$\gamma_1 = 0$, and has length of order $2\omega/V$ (Figure \ref{fig:kperp}). 
The minor axis of the ellipse is along $\hat L_{1,\perp}$ (which is orthogonal to the line $\gamma_1 =0$) and has width of order $2/L$. 
   
\subsection{\label{sec:twoarm}Differential arm case}
The differential length fluctuations $\delta L$ can then be expanded into the length fluctuations in the individual arms plus cross-terms that represent the correlation between the arms: 
\be\label{expand}
\,\big\langle \tilde{\delta L^*}(\omega) \tilde{\delta L}(\omega') \big\rangle  = \,\big\langle \tilde{\delta L^*_1}(\omega) \tilde{\delta L_1}(\omega') \big\rangle 
+ \,\big\langle \tilde{\delta L^*_2}(\omega) \tilde{\delta L_2}(\omega') \big\rangle \\   
- \,\big\langle \tilde{\delta L^*_1}(\omega) \tilde{\delta L_2}(\omega') \big\rangle  - \,\big\langle \tilde{\delta L^*_2}(\omega) \tilde{\delta L_1}(\omega') \big\rangle
\ee

\begin{figure}[htbp]
\begin{center}
\includegraphics[width=10cm]{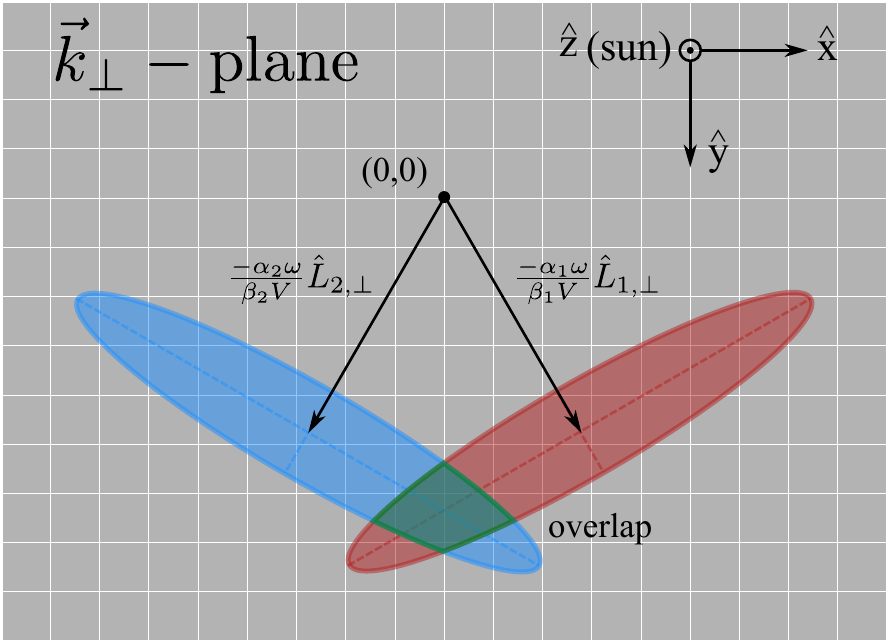}
\caption{This figure is a cartoon illustrating the regions in the $k_{\perp}$-plane that dominate various integrals. The  integral in Eq.~(\ref{LL1}), for
$\big\langle \tilde{\delta L^*_1}(\omega) \tilde{\delta L_1}(\omega') \big\rangle$ is dominated by the red elliptical region centered at $\frac{-\alpha_1 \omega}{\beta_1 V} \hat L_{1,\perp}$, with semi-minor axis parallel to $\hat L_{1,\perp}$ and of  length $\sim \pi/L$, and with semi-major axis liying along the line $\gamma_1 = 0$, and having length $\sim \omega/V$. Likewise, the blue elliptical region in $\vec k_{\perp}$-space dominates the contribution of the corresponding integral for $\delta L_2$. The red and blue ellipses depict the regions in which either $\gamma_1$ or $\gamma_2$ is small. The green shaded area depicts the region where both $\gamma_1$ and $\gamma_2$ are small. The integral for the cross-term  $\langle \delta L_1  \delta L_2 \rangle$  is dominated (roughly) by the contributions from the green shaded area.  Intuitively, the fact that the green region is much smaller than either the red or blue regions is what allows us to neglect the cross terms, compared to the first two terms, in Eq.~(A20).
}
\label{fig:kperp}
\end{center}
\end{figure}

The first two terms are described by (\ref{final}) and are identical except that $\beta_1 \rightarrow \beta_2$ for the second term.  By basically retracing the
calculation that led to Eq.~(\ref{term1}), one finds that last two terms in Eq.~(\ref{expand}) sum to 
\be\label{termsum}
\left(\chi L\right)^{2}  \frac{P_0 k_0^{11/3}}{V} \, \int \frac{d^2\vec k_{\perp}}{2\pi} \sinc\frac{\gamma_1 L}{2} \sinc\frac{\gamma_2 L}{2} \cos \frac{(\gamma_1 -\gamma_2)L}{2} \, \delta(\omega - {\omega}') \, \left( \left(\frac{\omega}{V}\right)^2 + k_{\perp}^2\right)^{-11/6} ;
\ee
i.e., the same expression as (\ref{term1}), but with $\sinc^2(\gamma_1 L/2)$ replaced by the $ -2 \sinc(\gamma_1 L/2) \sinc(\gamma_2 L/2) \cos( (\gamma_1 -\gamma_2)L/2 )$.
For the case $\omega \gg \omega_0$ (i.e., for the frequency region of most interest), the maximum of the integrand for cross-terms is comparable to that for Eq.~\ref{term1} 
 but the integrand remains of order that value only in a region where \emph{both} $\gamma_1$ \emph{and} $\gamma_2$ are small.  For a  $60^{\circ}$ angle between the two arms, this region is much smaller than the region where \emph{either} $\gamma_1$ \emph{or} $\gamma_2$ is small (Figure \ref{fig:kperp}). More precisely, it is straightforward to show that the magnitude of the cross-terms is smaller than the first two terms on the right-hand-side of (\ref{expand}) by a factor of order $\omega_0/\omega$. Thus we can neglect the cross-terms and the noise power spectrum for a 2-armed LISA is approximately twice the noise power spectrum for a single arm.  
 
\section{How the above results are affected by anisotropy of the Kolmogorov spectrum}
So far in this paper, we have modeled the electron-density fluctuations as having an \emph{isotropic} Kolmogorov spectrum. 
However, theoretical arguments, confirmed by observations, show that when the eddy size $d$ shrinks below the value where the fluid velocity difference across the eddy exceeds the Alfven speed, then the turbulent cascade of energy to lower lengthscales becomes less efficient for wave vectors parallel to $\vec B$ than for wave vectors perpendicular to $\vec B$ \cite{Lithwick_Goldreich_Sridhar_2007}.  In the solar wind at $1 AU$, both simulations and measurements show that this transition to anisotropic turbulence occurs at wave vector
$k_B \approx 1.2 \times 10^{-5}\,\mathrm{km}^{-1}$.  For satellites moving with at velocity V with respect respect to the wind, this transition lengthscale corresponds to frequency $\omega_B \approx V k_B \approx 5 \times 10^{-3}\,\mathrm{rad/s}$ (for $V = 400\,\mathrm{km/s}$) or $f_B \approx 0.8\,\mathrm{mHz}$ ~\cite{Chen2016}.  In position space, it is useful to picture the turbulent eddies becoming more elongated in the $\hat B$ direction as one goes to length scales below $1/k_B$ (even though this fluid-like picture 
is not really physically appropriate for the solar wind, which is better described by nearly collision-less MHD).
More precisely, let  $\vec k = k_B \hat B + \vec k_{\perp B}$, and define $k_{\perp B} \equiv |\vec k_{\perp B}|$, and define
$L_B \equiv {k_B}^{-1}$.
In a series of papers of increasing generality (see \cite{Goldreich2001}, \cite{Maron_Goldreich_2001}, \cite{Lithwick_Goldreich_Sridhar_2007},  and references therein), Goldreich and collaborators showed that the anisotropic spectrum at scales $k_{\perp B} \agt 1/L_B$ the factor $P_0 k_0^{11/3}  |\vec k|^{-11/3}$ in Eq.~(\ref{eq:ne3d}) transitions to the following form.

\be\label{ani_spectrum}
P_0 k_0^{11/3} \times 
           (L_B)^{1/3} k_{\perp B}^{-10/3} \Theta\left((L_B)^{-1/3} \, k_{\perp B}^{2/3} -  |k_B| \right)                             
\ee

The conditions under which this holds were extended to the conditions in the solar wind in \cite{Lithwick_Goldreich_Sridhar_2007}.
Note that approximating the spectrum with a $\Theta$ function in this way--and so making the spectrum drop discontinuously to zero as $k_{\perp B}$ drops from slightly above $|k_B|^{3/2} {L_B}^{1/2}$ was  to slightly below -- is clearly unphysical, and not meant to be taken to completely literally.  In \cite{Lithwick_Goldreich_Sridhar_2007},  it is derived using extremely insightful order-of-magnitude calculations, which by themselves, however, are not sufficient to describe the spectrum in finer detail.
When we use Eq.~(\ref{ani_spectrum}) to calculate the spectrum of $\delta L$ below, our results will generally be discontinuous with our isotropic results at $\omega = \omega_B$ (though they will match each other at $\omega_B$ to within a factor of order one). Under these circumstances, it is physically reasonable to slightly modify our anisotropic results in order to enforce continuity at $\omega = \omega_B$. However we shall do this only at the final step, so that the reader can first see the answer if one were to take Eq.~(\ref{ani_spectrum}) completely literally.

Regarding both radar ranging and in-situ measurements, the "lore" in the literature is that this B-induced anisotropy has only relatively small effect on the observed power spectrum. To our knowledge, there is no publication in which this lore is theoretically demonstrated--either by analytic calculations or results of numerical simulations. The following physical picture is useful for understanding. (We found this analogy while doing background reading, but were unable to re-locate it, so apologize for not being able to cite it here.) Picture a bundle of eddies elongated in the $\hat B$-direction similar to a box of pencils. Then picture the path of the in-situ satellite, or the radar ranging beam, as a line passing through the pencil box.  For generic orientations of the pencil box and the line, the line passes through each pencil in approximately the short direction, i.e., for typical orientations, the length of the pencils makes little difference.  Similarly, integrated $n_e$ fluctuations along generic directions in the solar wind  should be roughly the same as for an isotropic Kolmogorov spectrum.

While the "pencil argument" appears to apply to the LISA case too, in the following we shall show explicitly that the B-induced anisotropy has only a very modest effect on our above results for
the solar wind contribution to LISA's noise spectral density. We begin by restricting ourselves to cases where $\vec V \perp \vec L$.  This should not impact our results very much, since $\vec V$ is alway roughly orthogonal  to $\vec L$ (deviating by a maximum of $\pi/6$ radians, and more typically by $\sim \pi/12$.  As throughout this paper, we take $\vec V$ to be along the $\hat z$-axis; additionally, in this Appendix, we take $\vec L$ to be along the $\hat y$-axis.

Starting with the above simplification, we consider in the next three subsections three special cases for the direction of $\vec B$.  By the end, it should be clear that these special cases should bound the ratio 
${S_L^\text{aniso}}(f)/{S_L^\text{iso}}(f)$. And we shall see that even for the most extreme cases, $\vec B$ along $\vec V$ and  
$\vec B$ along $\vec L$, that the anisotropic spectrum changes our result for the noise amplitude by less than a factor 2 in the frequency range of interest. In a further subsection, we will argue that the differential two-arm result also changes very little when we include anisotropic effects.

\subsection{$ \vec B$ along the x-axis}
First recall that under our conventions, $\vec B$ along $\hat x$ means that $\vec B$ is orthogonal to both $\vec V$ and $\vec L$.   So the pencil argument, to the extent that it is valid at all, should certainly apply in this case.   We shall not go through all the steps in the following calculations; instead we will point out how the calculation changes from the isotropic case
we solved in Appendix A. For the anisotropic case, the rhs of Eq.~(\ref{term1}) gets replaced by
\be\label{ani_term1_x}
\left(\chi L\right)^{2}  \frac{P_0 k_0^{11/3}}{V} \, {L_B}^{1/3} \int \frac{dk_x\, dk_y }{(2\pi)^2} \left( (\omega/V)^2 + {k_y}^2\big)^{-10/6} \Theta\big({L_b}^{-1/3}(\omega/V)^2 + {k_y}^2\right)^{1/3} - |k_x|\big) \sinc^2(k_y L/2) (2\pi) \delta(\omega - {\omega}')
\ee
Since $k_x$ does not appear in the integrand, except for inside the argument of the $\Theta$ function, the integral over $k_x$ is completely trivial, and Eq.~(\ref{ani_term1_x}) reduces to  
\be\label{ani_x2}
\left(\chi L\right)^2 \frac{P_0 k_0^{11/3}}{V} \, {L_B}^{1/3} 2 \int \frac{dk_y }{2 \pi} {L_b}^{-1/3}\left( (\omega/V)^2 + {k_y}^2 \right)^{-4/3}  \sinc^2(k_y L/2) \delta(\omega - {\omega}')
\ee
As in the isotropic case, the $\sinc^{2}$ means that (in the LISA band) the integral is dominated by the region where $|k_y| \alt 2\pi/L$, and so, after a bit of algebra, 
(\ref{ani_x2}) becomes

\bea\label{ani_x3}
& \ &\left(\chi\,L\right)^{2}\,\left[\frac{3}{5}  \,P_0\, k_0^{11/3}\, V^{2/3} |\omega|^{-5/3}\right] \big(\frac{10}{3}\big) \, (\frac{V}{L\omega}) \delta(\omega - {\omega}') \, \\
& \approx &\left(\chi\,L\right)^{2}\,\left[\frac{3}{5} \,P_0\, k_0^{11/3}\, V^{2/3} |\omega|^{-5/3}\right] \big(\frac{25}{9}\big) \, (\frac{V}{L\omega}) \delta(\omega - {\omega}') 
\eea
where in the last line we changed $10/3$ to $25/9$ by hand (a $17\%$ change) to enforce continuity with 
So for $\vec B$ along $\hat x$, our best estimate shows no substantial difference from the isotropic case. 

\subsection{$\hat B$ along $z$-axis}
This is the case where $\hat B$ is parallel (or anti-parallel) to the wind velocity $\vec V$.  For this case, the rhs of Eq.~(\ref{term1}) becomes
\be\label{ani_term1_z}
\left(\chi L\right)^2  \frac{P_0 k_0^{11/3}}{V} \, {L_B}^{1/3} \int \frac{dk_x\, dk_y}{(2\pi)^2} \big( {k_x}^2 + {k_y}^2\big)^{-10/6} \Theta\big({L_b}^{-1/3} ({k_x}^2 + {k_y}^2\big)^{1/3} - \omega/V \big) sinc^2(k_y L/2) (2\pi) \delta(\omega - {\omega}')
\ee
Again, the $\sinc^2(k_y L/2)$ term guarantees that the integral is dominated by the region where $|k_y|$ is small, we can approximate the above by 
\bea\label{ani_term1_z2}
&\ &\left(\chi L\right)^2  \frac{P_0 k_0^{11/3}}{V} \, {L_B}^{1/3} (L)^{-1} \int_{\infty}^{\infty}  d k_x \, |k_x|^{-10/3}  \Theta\big({L_b}^{-1/3} (|k_x|^{2/3}  - \omega/V \big) \delta(\omega - {\omega}') \\
&=& (\chi L)^2 \frac{P_0 k_0^{11/3}}{V} \, {L_B}^{1/3} (L)^{-1} 2 \int_{k_{x,min}}^{\infty} (k_x)^{-10/3} d k_x \,   \delta(\omega - {\omega}')
\eea
where (from the $\Theta$ function) $k_{x,min} = L_b^{1/2} (\omega/V)^{3/2} $, giving

\bea\label{ani_term1_z2}
&\ &\ (\chi L)^2  \frac{P_0 k_0^{11/3}}{V} \, {L_B}^{1/3} (L)^{-1} \frac{6}{7}  \big(L_b^{1/2} (\omega/V)^{3/2}\big)^{-7/3} \delta(\omega - {\omega}') \\
&=&\left(\chi\,L\right)^{2}\,\big[\frac{3}{5} \,P_0\, k_0^{11/3}\, V^{2/3} |{\omega}|^{-5/3}\big] \big(\frac{10}{7}\big)\ (\frac{V}{L\omega}) F(\omega)\delta(\omega - {\omega}') \\
& \approx & \left(\chi\,L\right)^{2}\,\big[\frac{3}{5} \,P_0\, k_0^{11/3}\, V^{2/3} |{\omega}|^{-5/3}\big] \big(\frac{25}{9}\big)\ (\frac{V}{L\omega}) F(\omega)\delta(\omega - {\omega}') 
\eea
where in the last line we adjusted $10/7$ to $25/9$ by hand to maintain continuity at $\omega_B \equiv V/L_B$, and where $F(\omega)$ is defined as
\be\label{defF}
F(\omega) \equiv \begin{cases}
           \omega/\omega_B \ \ \ \ \ \ \ \ \  \text{for $\omega > \omega_B $} \\
           1 \ \ \ \ \ \ \ \ \ \ \ \ \ \ \ \  \text{otherwise}
           \end{cases}   
\ee                  
Thus when $B$ is aligned with $V$,  the spectral density of the noise for $f > f_B$ is reduced compared to the isotropic case.
Since the $B$-field impedes the cascade of turbulent power along the $\hat B$ direction, that accords with our expectations.

\subsection{B along y-axis}
This is the case where $\vec B$ is parallel to the arm $\vec L$.  In this case,
the rhs of Eq.~(\ref{term1}) becomes
\be\label{ani_term1_y}
 \left(\chi L\right)^2 \frac{P_0 k_0^{11/3}}{V} \, {L_B}^{1/3} \int \frac{dk_x\, dk_y}{(2\pi)^2} \left( (\omega/V)^2 + {k_x}^2\right)^{-10/6} \Theta\left({L_b}^{-1/3}(\omega/V)^2 + {k_x}^2)^{1/3} - |k_y|\right) \sinc^2(k_y L/2) (2\pi) \delta(\omega - {\omega}')
\ee
Again, because of the $\sinc^2$ term, the integral is dominated by the region $k_y \alt 2\pi/L$. In this region,
it is easy to show that the argument of the $\Theta$ function is positive if $(L \omega/V) > 2^{3/2}(L_B/L)^{1/2}$,
which clearly is always satisfied in the LISA frequency band. Thus we can replace the $\Theta$ function by
one.  Then the integral over $k_y$ is again trivial,
leaving us with
 \be
\left(\chi L\right)^2 \frac{P_0 k_0^{11/3}}{V} \, {L_B}^{1/3} (L)^{-1} \int_{-\infty}^{\infty} dk_x  \big( (\omega/V)^2 + {k_x}^2\big)^{-10/6}  \delta(\omega - {\omega}') \ 
\ee
Using
\be 
\int_{-\infty}^{\infty} dk_x  \left( (\omega/V)^2 + {k_x}^2\right)^{-10/6} \approx (9/5)(\omega/V)^{-7/3}
\ee
the above is easily shown to equal
\bea
&\ &(\chi L)^2  \frac{P_0 k_0^{11/3}}{V} \, {L_B}^{1/3} L^{-1} (\frac{9}{5})(\omega/V)^{-7/3}  \delta(\omega - {\omega}') \\
&=&\left(\chi\,L\right)^{2}\,\left[\frac{3}{5} \,P_0\, k_0^{11/3}\, V^{2/3} |\omega|^{-5/3}\right] 3 {\beta_1}^{5/3}  (\frac{V}{L\omega}) 
F^{-1}(\omega) \delta(\omega - {\omega}') \, \\
& \approx &\left(\chi\,L\right)^{2}\, \left[\frac{3}{5} \,P_0\, k_0^{11/3}\, V^{2/3} |\omega|^{-5/3}\right] \frac{25/9} {\beta_1}^{5/3}  (\frac{V}{L\omega}) 
F^{-1}(\omega) \delta(\omega - {\omega}') \, \\
\eea
where we defined $F(\omega)$ above in Eq.~(\ref{defF}), and where in last line we adjusted $3 \rightarrow \frac{25}{9}$
to maintain continuity with $\omega < \omega_B$

The factor $F(\omega)$ indicates that for $\omega > \omega_B$, the one-arm power can be somewhat above the isotopic case, but only very mildly so. E.g., as the frequency $f$ increases from $f_B = 0.8$mHz to $50$mHz,  the factor
$(\frac{L_B\omega}{V})^{1/3}$ in the noise increases by only $4$. 
Compared to isotropic case, this is only a factor $2$ in noise amplitude.

\subsection{Summary and implications of last three subsections}
We have seen that the B-field-induced anisotropy of the turbulence spectrum only affects the spectrum above $f_B \approx 0.8\,\text{mHz}$. For the case $\hat B$ along $\hat x$, the spectrum remains the same as for the isotropic case.
For the case of $\hat B$ parallel to $\hat z$ noise amplitude (the square root of the noise power derived above) in the anisotropic case is the isotropic result times $(f_B/f)^{1/6}$, while for $\hat B$ parallel to $\hat y$, the correction factor is $(f/f_B)^{1/6}$. Any other B-field direction should interpolate between these two correction factors, and so make less than a factor of difference 2 in noise amplitude up to $f= 50\,\text{mHz}$. 
 
\subsection{Anisotropy and the differential 2-arm case}
In Appendix A, we showed that 2-arm differential solar-wind-noise spectrum is basically the sum of the one-arm spectra, and that those two spectra are almost the same -- differing only by the geometrical term that depends on $\alpha, \beta$.
That followed from showing that the cross-terms in Eq.~(\ref{expand}) are small compared to each of the one-armed terms.
What can change in the anisotropic case?  First the spectra for $\delta L_1$ and $\delta L_2$ can differ somewhat more,
since they now also depend on the angle between the arm and the B-field.  But we have shown that change is also generally quite modest.  What about the cross-terms? Their relative size could possibly increase somewhat whenever $\hat B$ is roughly aligned with $\vec L_1 - \vec L_2$.  But we have not bothered to calculate that effect, for the simple reason that any
increased correlation between the two arms can only {\it decrease} the differential signal.  That is, the "likely small" effect
of this alignment would only decrease the noise in the two-arm differential signal, rendering LISA's solar-wind noise even less important.  

\section{Comparison with other analyses}
The analysis in this paper inspired by a similar analysis by Smetana~\cite{smetanaMNRAS} which also used \emph{Wind}/SWE data to estimate the solar plasma displacement noise effect for LISA. That analysis utilized 14 representative days of "normal" solar activity as well as 6 days corresponding to solar events. All 20 of these days are included in the data set we presented in section \ref{sec:data}.  The general procedure for both analyses consists of the same two steps: estimating the power spectral density from \emph{Wind}/SWE timeseries data and then computing the resulting effect in LISA using a transfer function model.  Contrary to our analysis, the analysis by Smetana indicated that the solar plasma effect would exceed the allocated LISA noise levels, leading to a degradation in sensitivity to gravitational waves. Part of the difference in our analyses comes from the difference in transfer functions between the in-situ point measurements of \emph{Wind}/SWE and the column-density effect in LISA. To assess the impact of this transfer function difference, we computed the effect using the transfer function of Smetana, which assumes that the entire LISA arm experiences electron density fluctuations that are identical in amplitude and matched in phase. In such a case, the relationship between the in-situ electron density fluctuations and the induced path-length fluctuations is simply $\delta L_1(t) = \chi L\delta N_e(t)$.  The power spectrum of apparent length fluctuations is then related to the power spectrum of electron density fluctuations for an in-situ measurement by the coupling factor $(L\,\chi)^2\approx 1.61\times10^{-20}\,\textrm{cm}^8$.  Compared with the transfer function accounting for LISA's orientation and the geometry of the solar wind, this overly-simple transfer function is missing a $f^{-1}$ filtering effect for Fourier frequencies $f \gtrsim 25\,\mu$Hz.

\begin{figure}[htbp]
\begin{center}
\includegraphics[width=14cm]{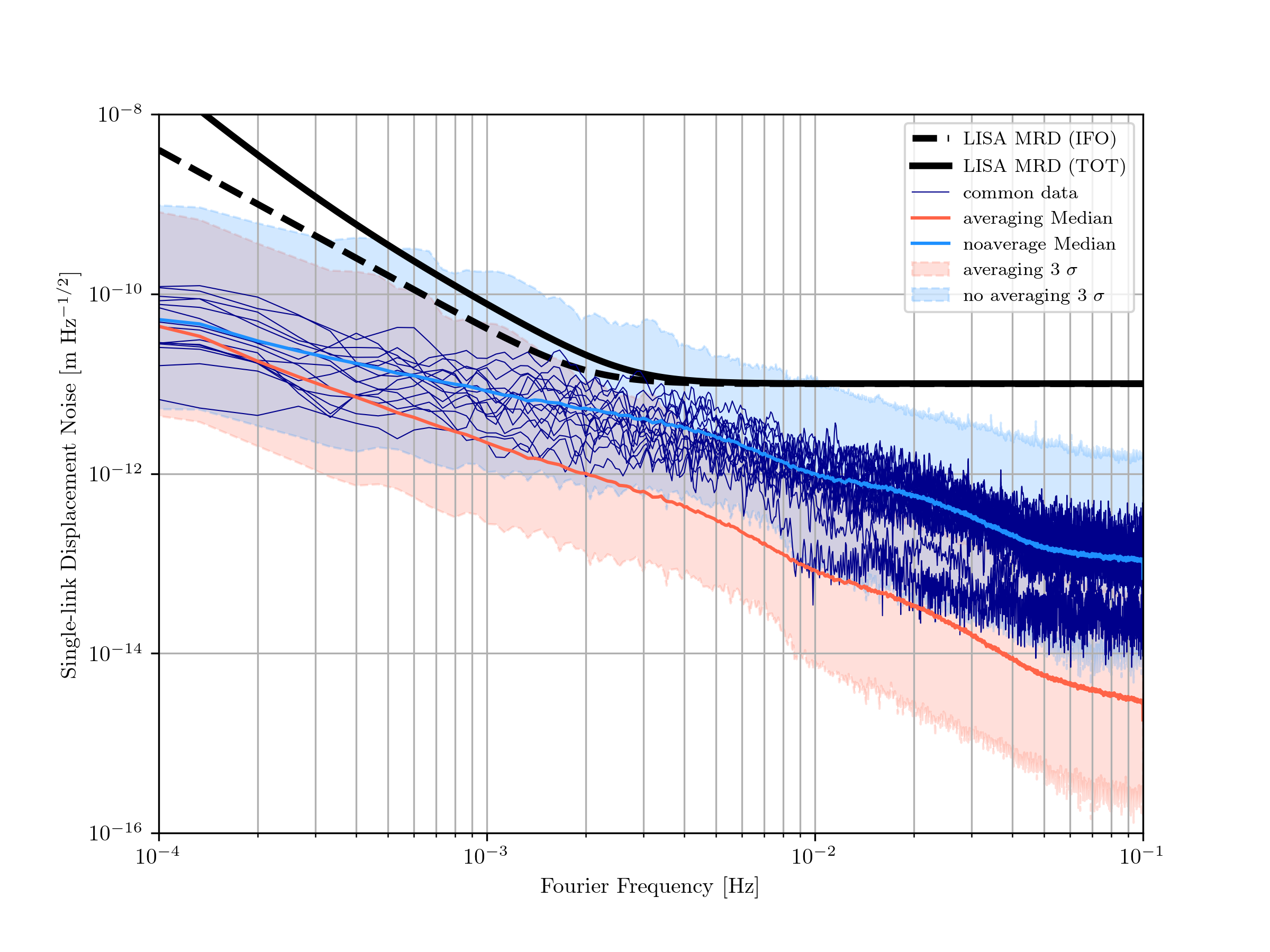}
\caption{Comparison of the effective single-link displacement noise induced by solar plasma for the case of the single-arm transfer function derived in (\ref{eq:L1arm_approx}) and the (non-physical) case where the fluctuations in electron density are identical and coherent along the entire LISA arm.  In both cases, the electron density fluctuations are taken from the \emph{Wind}/SWE data as described in (\ref{sec:data}). Note that, even for the non-physical case, while portions of the 3-$\sigma$ region do exceed the overall LISA noise allocation, the median displacement noise lies below the allowed contributions from other noise sources, as do the thin solid curves which correspond to 13 of the particular data epochs utilized in \cite{smetanaMNRAS}. The single-arm averaging case is identical to the curves in Figure \ref{fig:1arm} and represents our best estimate for the effect.}
\label{fig:compare}
\label{default}
\end{center}
\end{figure}

Figure \ref{fig:compare} compares the two transfer functions using our ensemble of spectra. The effect of the averaging in our transfer function is clear in the steeper slopes, and lower amplitudes, in the LISA band.  However, even the (non-physical) transfer function without this averaging effect still estimates the median spectra to be below the allocation for other noise sources in LISA. The 3-$\sigma$ region does extend above the allocation, at a roughly similar level to the curves presented in Smetana's analysis.  The solid curves correspond to the 14 "normal" days used in Smetana's analysis, all of which are below the LISA allocations.  This suggests that an additional reason for the difference in conclusions lies in the estimation of the electron density power spectra from the corresponding timeseries. 

\begin{acknowledgments}
The authors would like to thank Adam Smetana for his paper and presentation at the 2020 LISA Symposium on this same topic which inspired this analysis as well as for some helpful discussions during and after the LISA Symposium. Part of the work was carried out at the Jet Propulsion Laboratory, California Institute of Technology,  under contract with the National Aeronautics and Space Administration.
Additionally, we would like to thank Lynn Wilson for his helpful comments.
\end{acknowledgments}

\bibliography{solarWindLISA}

\end{document}